\documentclass[conference]{IEEEtran}
\usepackage{cite}
\usepackage{amsmath,amssymb,amsfonts}
\usepackage{tabularx}
\usepackage{algorithmic}
\usepackage{graphicx}
\graphicspath{ {./Figures/} }
\usepackage{textcomp}
\usepackage{xcolor}
\usepackage{xspace}
\usepackage{pgfplots}
\pgfplotsset{compat=1.18}
\usepackage{tikz}
\def\BibTeX{{\rm B\kern-.05em{\sc i\kern-.025em b}\kern-.08em
    T\kern-.1667em\lower.7ex\hbox{E}\kern-.125emX}}
\newcolumntype{C}{>{\centering\arraybackslash}X}

\begin{document}

\title{Efficiently Reproducing Distributed Workflows in Notebook-based Systems}

 \author{
\IEEEauthorblockN{
 Talha Azaz\IEEEauthorrefmark{1},
 Raza Ahmad\IEEEauthorrefmark{2},
 Md Saiful Islam\IEEEauthorrefmark{3},
 Douglas Thain\IEEEauthorrefmark{3},
 Tanu Malik\IEEEauthorrefmark{1}}\\

\IEEEauthorblockA{
\textit{\IEEEauthorrefmark{1}University of Missouri, Columbia, MO, USA;}
\textit{\IEEEauthorrefmark{2}DePaul University, Chicago, IL, USA;}\\
\textit{\IEEEauthorrefmark{3}University of Notre Dame, Notre Dame, IN, USA;}\\
\text{tay2f, tanu@missouri.edu},
\text{raza.ahmad@depaul.edu},
\text{mislam5, dthain@nd.edu}
}}


\newcommand{\toolname}{{\texttt{NBRewind}}\xspace}

\maketitle

\begin{abstract}
Notebooks provide an author-friendly environment for iterative development, modular execution, and easy sharing. Distributed workflows are increasingly being authored and executed in notebooks, yet sharing and reproducing them remains challenging. Even small code or parameter changes often force full end-to-end re-execution of the distributed workflow, limiting iterative development for such workloads. Current methods for improving notebook execution operate on single-node workflows, while optimization techniques for distributed workflows typically sacrifice reproducibility. We introduce NBRewind, a notebook kernel system for efficient, reproducible execution of distributed workflows in notebooks. NBRewind consists of two kernels—audit and repeat. The audit kernel performs incremental, cell-level checkpointing to avoid unnecessary re-runs; repeat reconstructs checkpoints and enables partial re-execution including notebook cells that manage distributed workflow. Both kernel methods are based on data-flow analysis across cells. We show how checkpoints and logs when packaged as part of standardized notebook specification improve sharing and reproducibility. Using real-world case studies we show that creating incremental checkpoints adds minimal overhead and enables portable, cross-site reproducibility of notebook-based distributed workflows on HPC systems.
\end{abstract}



\section{Introduction}
\label{sec:Introduction}

\begin{figure}[ht]
  \centering
  \includegraphics[width=0.85\linewidth]{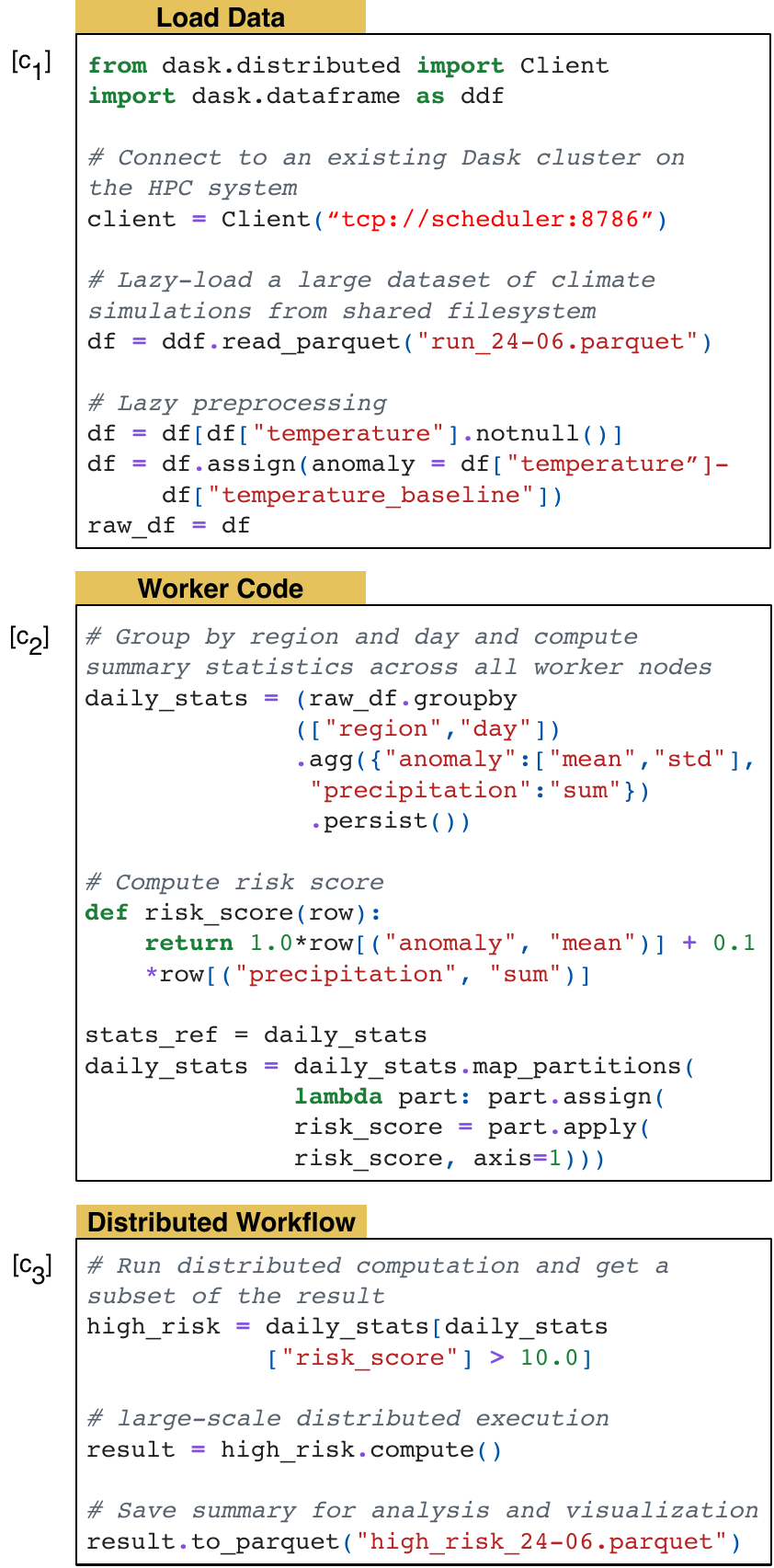}
  \caption{Example of a distributed workflow in a notebook with a manager creating worker threads to compute summary statistics.}
  \label{fig:notebook-example}
\end{figure}

Notebooks~\cite{kluyver2016jupyter}  provide an author-friendly environment that enables iterative development, modular cell-level execution, and easy dissemination of computational analyses across scientific disciplines\cite{mendez2019toward,bascunana2023impact,castilla2023jupyter,choi2023comparing}. Increasingly, distributed scientific workflows are prototyped and executed within notebook interfaces. For example, users may develop a notebook using libraries like Dask~\cite{Rocklin2015DaskPC} (as shown in Figure~\ref{fig:notebook-example}),  Parsl~\cite{babuji2019parsl} or TaskVine~\cite{taskvine-works-2023}, which enable specification of the distributed workflow from within notebooks. While such notebooks are often helpful for iterative development, they are difficult to share and reproduce across different cluster environments---often essential in collaborative scientific settings~\cite{islam2025backpacks}.

A core strength of notebooks is the ability to efficiently explore variations in parameters or code while preserving consistency of results. Consider the distributed workflow in Figure~\ref{fig:notebook-example}, where Cell 1 performs an expensive data-loading operation, Cell 2 defines a distributed map, group by, and aggregation, and a  \texttt{risk\_score} function to be computed over aggregated rows. Cell 3 triggers the execution of the distributed computation on multiple nodes using \texttt{compute}, producing a \texttt{parquet} file output. In a typical scenario, a scientist executes all cells, inspects the results, and then varies downstream parameters such as the  \texttt{risk\_score} threshold, for example, changing from 10.0 to 8.0. If the kernel remains alive and Dask has retained intermediate data in memory, the scientist may re-run only Cell 3.
However, the in-memory state of Dask is lost when the notebook is moved to another HPC cluster due to kernel restart, or when the notebook is shared with a collaborator. The collaborator inherits the notebook but none of its intermediate workflow state, forcing complete re-execution from raw data.

Although some systems offer coarse-grained checkpointing, existing notebook \textit{savepoints} are manual  and not designed for distributed workflows. Even if checkpointing is available, full checkpoints incur significant storage overhead, but more importantly, they do not determine which distributed computations must be partially re-executed when changes to notebook cells occur. 
In our running example, a checkpoint may allow skipping the expensive data-loading in Cell 1, but if the user modifies the \texttt{risk\_score} formula in Cell 2, both Cells 2 and 3 must be re-run. Rerunning entire Cell 3 is wasteful, since only part of Cell 3's distributed computation is logically affected by the edit --- the distributed group-by aggregation remains valid, and only the \texttt{risk\_score} computation should change. Current notebook systems have no mechanism to checkpoint distributed computation within a cell or to reuse previous distributed results while selectively re-executing only the modified portion.

In this paper, we address the dual challenge of repeating entire notebooks and entire distributed computation within a notebook by (i) checkpointing cell-level kernel state—including intermediate distributed workflow results, and (ii) selectively replaying only those tasks whose inputs or code have changed. The checkpoints of cells and the list of distributed tasks are created incrementally to save on checkpoint size. 


We present \toolname, a system consisting of two notebook kernels: audit and repeat. 
\toolname’s audit kernel performs automatic, \textit{incremental checkpointing}: as users execute each cell during iterative development, the kernel detects changes in the notebook’s kernel state and checkpoints only the variables and objects whose values have been updated relative to prior cells. This change detection is implemented by analyzing the variables in code and comparing them with other shared variables using a reverse index mapping of memory to variable names, efficiently determining which objects differ across cell boundaries.
Incremental checkpointing is essential for controlling storage cost, as naively checkpointing full kernel state after every cell leads to substantial space consumption. We show that without incremental checkpoints, users must provision prohibitively large storage volumes to support reproducibility.

For cells that contain distributed workflows, 
the audit kernel also interacts with the underlying distributed workflow engine 
These persisted objects are likewise indexed using a hash-based structure so that the system can determine which distributed tasks produce identical outputs and which must be invalidated when notebook code changes. 
When a distributed workflow notebook is shared and executed in a new cluster environment, \texttt{NBRewind}’s repeat kernel is used. The repeat kernel reconstructs the full state of any notebook cell by composing the previously saved incremental checkpoints. It then selectively replays distributed computation within that cell by determining which tasks are invalidated due to code or dataflow changes in the notebook. Only the invalidated tasks are re-executed; all other tasks reuse memoized intermediate results captured by the audit kernel. This enables partial re-execution of distributed workflows, ensuring that logically unaffected portions of a workflow are not recomputed.

The split-kernel design of \toolname builds on our prior work on containerizing dependencies for single-node notebooks~\cite{ahmad2025improving}. This design pattern has proven effective because it requires minimal changes to the notebook interface, preserves the interactivity of the notebook UI, and avoids intrusive modifications to user code. It is also compatible with emerging standardization efforts such as “backpack”-style artifacts that encapsulate notebook dependencies, data, and software environments~\cite{islam2025backpacks}. The audit and repeat kernels automatically provide the metadata and state snapshots required to construct such portable bundles.

Our primary contributions in this paper are as follows:

\begin{enumerate}
    \item Introducing incremental checkpointing and partial re-execution mechanisms for notebook cells that contain both single-node and distributed workflows.
    \item Designing a hash-based data structure for efficiently determining the minimal set of tasks that must be re-executed.
    \item Constructing a functional prototype of the audit and repeat kernels that implements automatic incremental checkpointing, distributed task memoization, and partial re-execution across notebook cells.
\end{enumerate}

The rest of the paper is organized as follows. Section~\ref{sec:Checkpointing in Notebooks} explains the mechanics of distributed computation in notebook workflows. Section~\ref{sec:NBRewind: Notebook Checkpointing and Partial Re-execution} gives an overview of the \toolname system and its architecture. Section~\ref{sec:Checkpointing in Notebooks} explains the checkpointing mechanism for saving local notebook state. Section~\ref{sec:Partial Re-Execution} details the process and methodology of partial re-execution in notebooks for distributed workflows. 
Section~\ref{sec:Experiments and Evaluation} highlights our experiments and evaluates the effectiveness of \toolname on our dataset of notebooks. We conclude in Section~\ref{sec:Conclusion} with a summary of our work and future directions.

\section{Distributed Workflows in Notebooks}
\label{sec:Distributed Workflows in Notebooks}

We describe relevant processes that maintain execution state of distributed workflows within a notebook.

\begin{figure}[t]
  \centering
  \includegraphics[width=\linewidth]{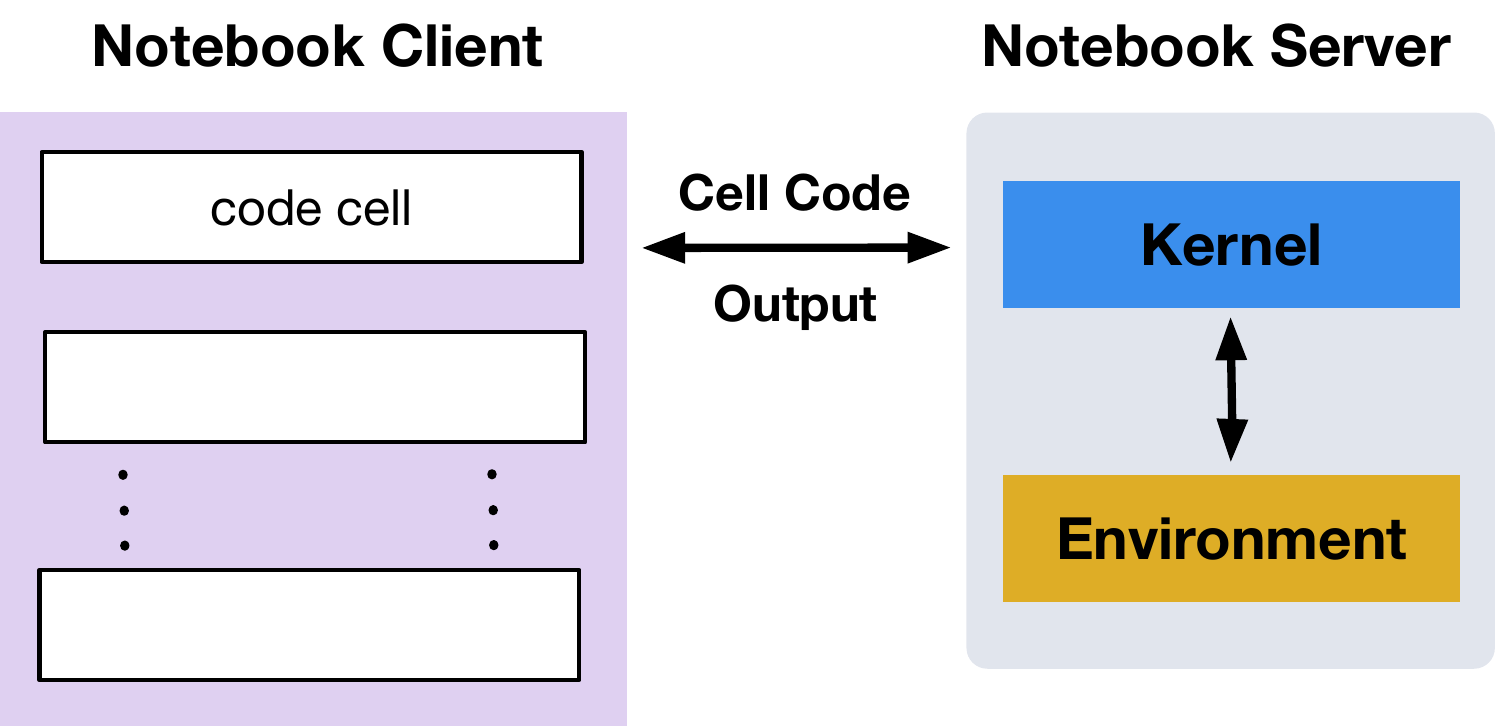}
  \caption{The notebook environment comprising web-based notebook client running the notebook code which interacts with the kernel process in the notebook server.}
  \label{fig:notebook-architecture}
\end{figure}

In general, notebooks are clients which users access and edit through a web browser. The code within each cell is executed on the server as part of a programming language-specific interactive \textit{kernel} process, as shown in Figure~\ref{fig:notebook-architecture}. Because the kernel persists across cells, the state produced by one cell (variables, objects, functions, configurations) becomes the starting point for subsequent cells. The output of the cell-level code evaluations from the kernel process is relayed to the notebook client. Since the kernel executes the user code, the kernel is also aware of the computational environment and the set of dependencies needed to run those code. Thus the notebook sends the user code one cell at a time, executes it using its configured environment, and returns the response of the executed code in the form of text and/or visualization, which is then displayed to the user through a web browser.

\begin{figure}[t]
  \centering
  \includegraphics[width=\linewidth]{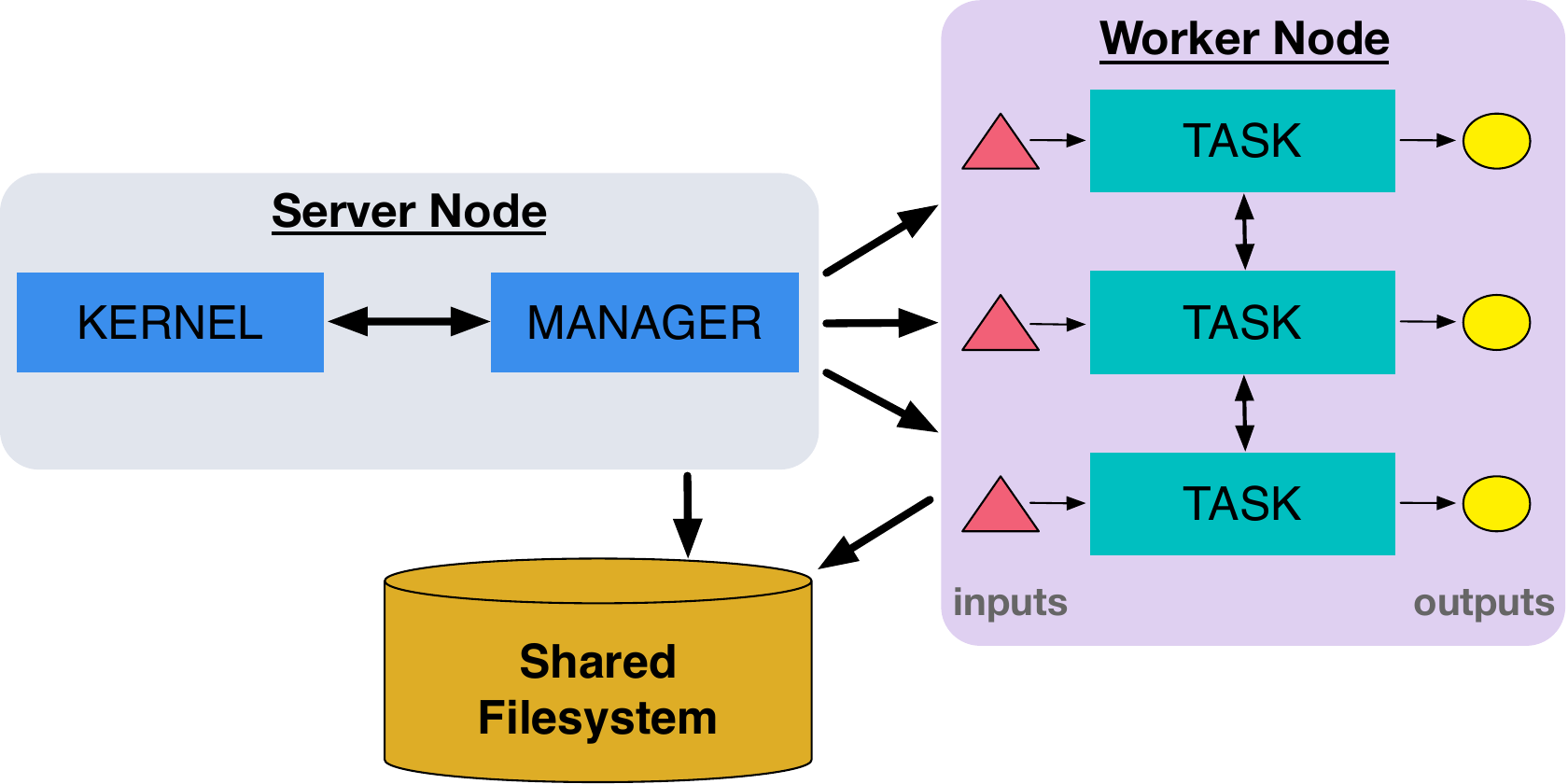}
  \caption{The distributed notebook workflow is illustrated when the kernel launches the manager which spawns several worker threads that run in parallel on the worker node.}
  \label{fig:distributed-workflow}
\end{figure}

When distributed workflows are developed in notebooks, one or more cells of the notebook may run some of their code either in a process spawned by the kernel or a remote node, giving rise to a distributed workflow within the execution life cycle of that cell. Typically a distributed workflow comprises of a manager and a collection of distributed workers. The manager instructs workers to run some tasks, which are essentially processes running in parallel. The tasks read and write files from a shared file system as shown in Figure~\ref{fig:distributed-workflow}. Depending on the distributed workflow system, the manager may organize tasks into a directed acyclic graph (DAG) structure to manage inputs and outputs and determine if a task is in progress or waiting for other tasks to finish. DAG-based management of tasks also helps in efficient resource utilization. 
To enable the distributed workflow within the notebook cell, users may use popular libraries such as TaskVine~\cite{taskvine-works-2023}, Dask~\cite{Rocklin2015DaskPC}, Parsl~\cite{babuji2019parsl}, which have Python extensions for use within notebooks.

The distributed workflow in notebooks  may span multiple cells. The kernel maintains local Python objects representing the workflow (e.g.,TaskVine task handles), while the manager process of the distributed system simultaneously maintains the remote state including partially completed tasks and cached intermediate results. Thus, the complete workflow state at any moment consists of both the kernel's local state and the distributed execution's state.

\section{NBRewind: Notebook Checkpointing and Partial Re-execution}
\label{sec:NBRewind: Notebook Checkpointing and Partial Re-execution}

\toolname is a system consisting of two notebook kernels: \textit{audit} and \textit{repeat}. Both kernels are wrappers around base kernels e.g., the Python or Xeus-SQL kernel. The user uses the audit kernel to execute the notebook using the language interpreter. As the audit kernel executes the notebook, it automatically checkpoints after every cell. This checkpointing allows a user to go back in time and use the state of the notebook at time \textit{t} instead of executing the other prerequisite cells from time $<$ \textit{t}. The audit kernel also intercepts the execution to determine all data and environment dependencies used by the notebook. 

When the notebook is shared along with the checkpoints, data and environment dependencies, the user re-runs the notebook using the repeat kernel. The repeat kernel determines if there is a change in the notebook code cell. If there is no change, the repeat kernel bypasses execution of the cell and returns the stored output. If the repeat kernel determines that the cell’s code has changed, this cell must be re-executed. For a changed cell, the repeat kernel reconstructs the checkpoint to execute the changed cell. Section~\ref{sec:Checkpointing in Notebooks} describes how to efficiently create and reconstruct checkpoints for each of the kernels respectively.






When a distributed workflow is run within the notebook, the user still uses \toolname’s audit and repeat kernels as described above. However during distributed execution, the audit kernel also determines if a cell issues any tasks. If so, the corresponding manager issuing the tasks is audited for task submissions. The audit kernel creates a log of all tasks that were submitted during execution. This log is available for sharing similar to checkpoints, and other data and environment dependencies. The repeat kernel uses the task log to optimize on re-execution when re-running the notebooks. Specifically, the repeat kernel bypasses execution of any task whose inputs have not changed. Section~\ref{sec:Partial Re-Execution} describes how to intercept the distributed workflow manager and audit tasks for partial re-execution of distributed notebook workflows.

\begin{figure*}[ht]
  \centering
  \includegraphics[width=0.9\linewidth]{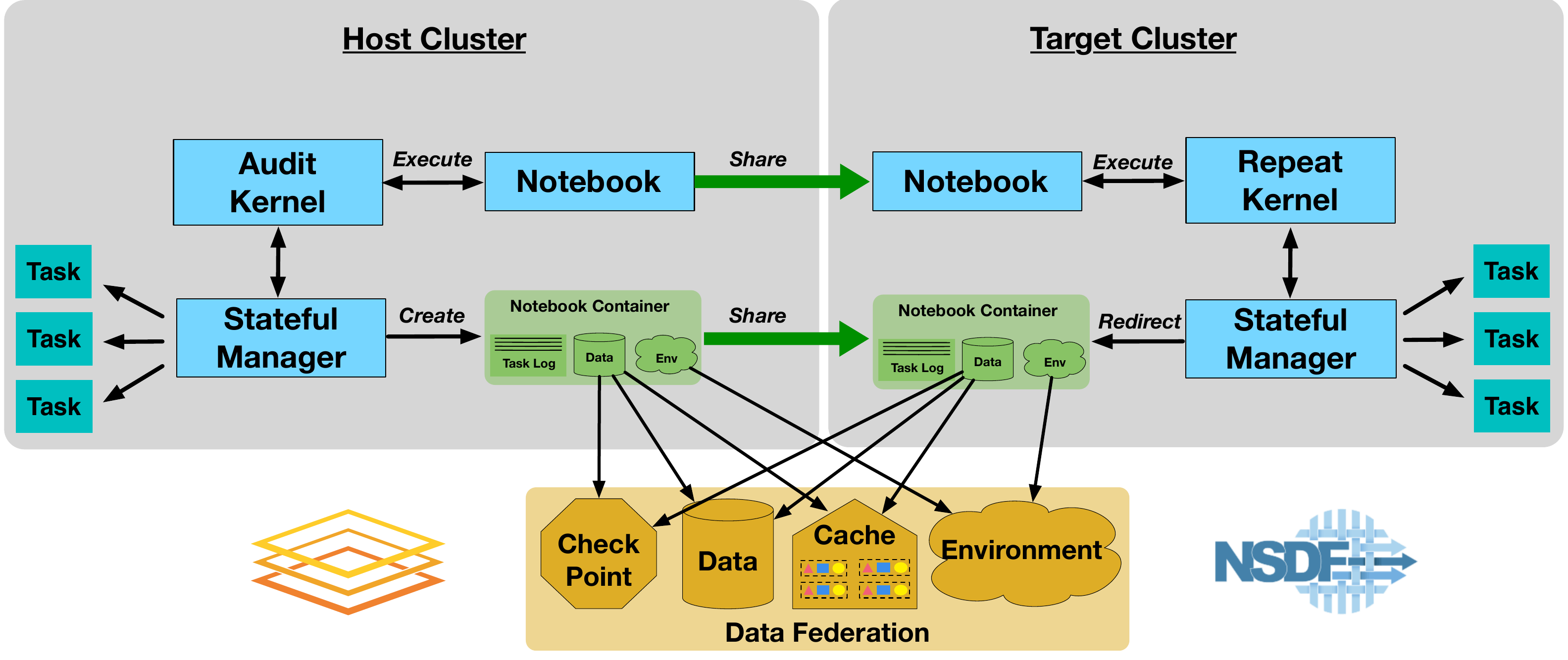}
  \caption{Architecture of \toolname is illustrated The audit kernel on the host cluster executes a notebook and creates its corresponding notebook container. The notebook and its container are shared with a collaborator to be executed in a target cluster using the repeat kernel. Large volumes of data in a notebook container may be fetched dynamically from a remote data store.}
  \label{fig:architecture}
\end{figure*}

The combined checkpointing of notebook cells and auditing of distributed tasks allows \toolname to improve scalability and reproducibility of notebooks, especially when users make changes and attempt to re-execute a shared notebook on a target cluster. Figure~\ref{fig:architecture} shows the notebook’s use of \toolname’s audit and repeat kernels to create checkpoints, task logs, input and intermediate data files and environment, which need to be containerized to make it shareable and reproducible across platforms. Our approach is to containerize the notebook program using application virtualization which captures all dependencies of the notebook by observing its execution~\cite{ahmad2025improving}. The audit phase generates a lightweight notebook container that includes all necessary dependency files to reproduce the notebook, while the repeat phase uses this container to seamlessly execute the corresponding notebook in the target environment. The outputs of the audit kernel can be standardized using recently proposed notebook backpacks~\cite{islam2025backpacks}. Similarly audited data for inputs, intermediate task outputs, environment files and checkpoints can be hosted either locally within the cluster or onto a data federation such as Open Science Data Federation~\cite{osdf-2024} or the National Science Data Fabric~\cite{nsdf-2024}. 

\section{Checkpointing in Notebooks}
\label{sec:Checkpointing in Notebooks}

The audit kernel creates a checkpoint after each cell execution. To create a checkpoint, the kernel could, in principle, capture its own state using an OS-level checkpointing mechanism such as Checkpoint/Restart in Userspace (CRIU), which snapshots the entire process, including memory, file descriptors, threads, and other OS-managed resources. While OS-level checkpointing is programming-language agnostic, it captures a substantial low-level state that is unnecessary for reproducing notebook execution and is expensive to store and restore.

Instead, the \toolname audit kernel performs application-level checkpointing of the Python session executing within the kernel process. Python maintains execution state in a global dictionary that maps names—such as variables, functions, and imported modules—to runtime objects. Naively persisting this dictionary after each cell execution is impractical due to its size and the high frequency of updates. To reduce checkpoint overhead, \toolname performs incremental checkpointing, persisting only those variables whose values may have changed as a result of executing a cell. As a first step, \toolname applies static notebook slicing similar to NBSlicer~\cite{shankar2022bolt}, by parsing the cell’s abstract syntax tree (AST) to conservatively identify variables that are read or written in the cell. This static analysis bounds the set of candidate variables that may affect the notebook state.



\begin{figure*}
  \centering
  \includegraphics[width=\linewidth]{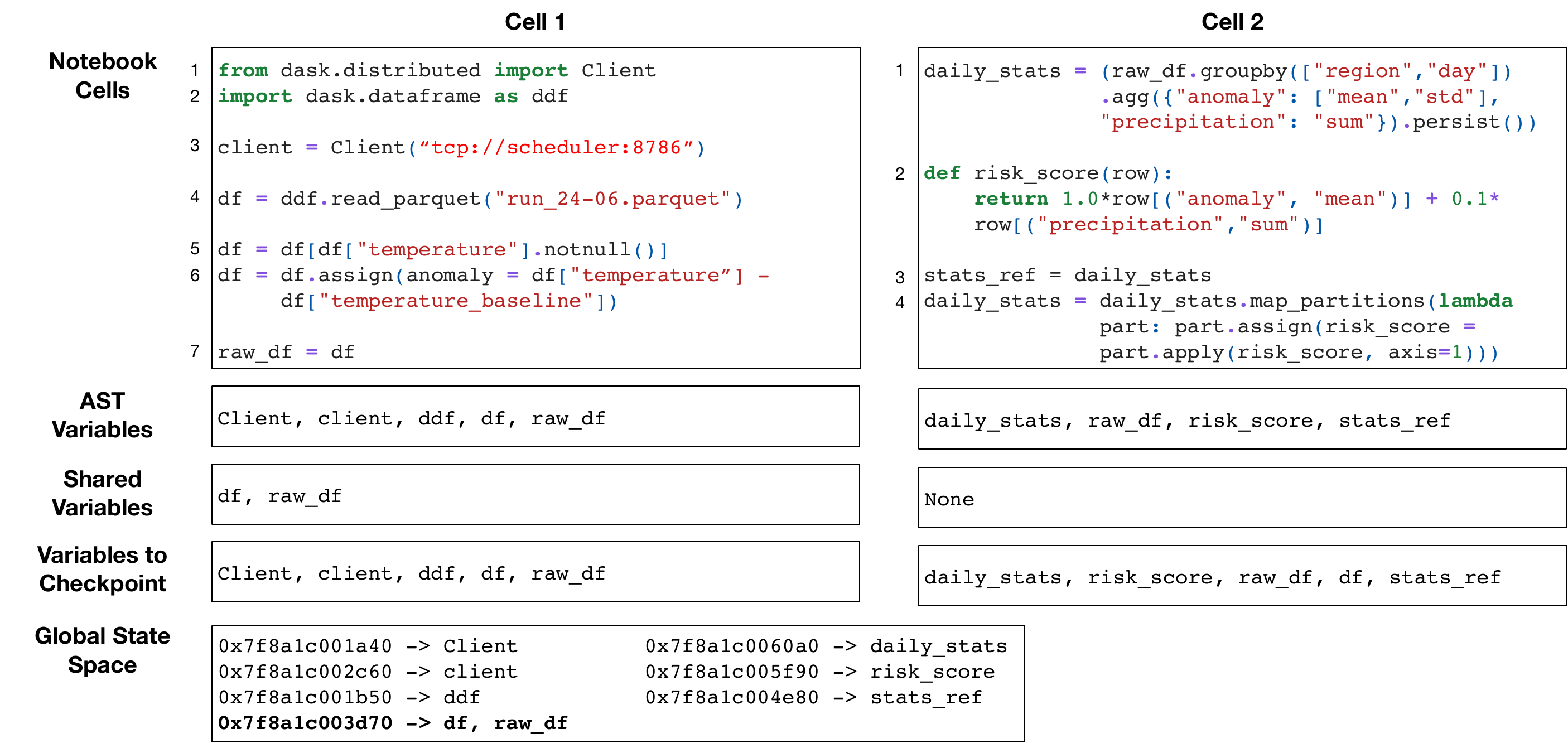}
  \caption{The internal states maintained during the \toolname workflow to create checkpoints, as illustrated by the first two cells of our example notebook.}
  \label{fig:nbrewind-workflow}
\end{figure*}

Static slicing alone, however, is insufficient to capture runtime dependencies arising from mutation and shared references. Therefore, \toolname augments static slicing with dynamic dependency tracking by observing object identities and shared memory references during execution. \toolname tracks shared variables, variables that reference the same underlying runtime object, and uses this information to determine additional variables that must be checkpointed transitively.

To support this process, \toolname maintains 
a reverse memory index, which maps memory references to all variables that alias them. Incremental checkpoints are constructed by combining statically identified variables with dynamically discovered shared dependencies. When a rollback is requested, the kernel reconstructs the notebook state by restoring the relevant subset of the globals dictionary from the corresponding incremental checkpoints.



\begin{figure}
  \centering
  \includegraphics[width=\linewidth]{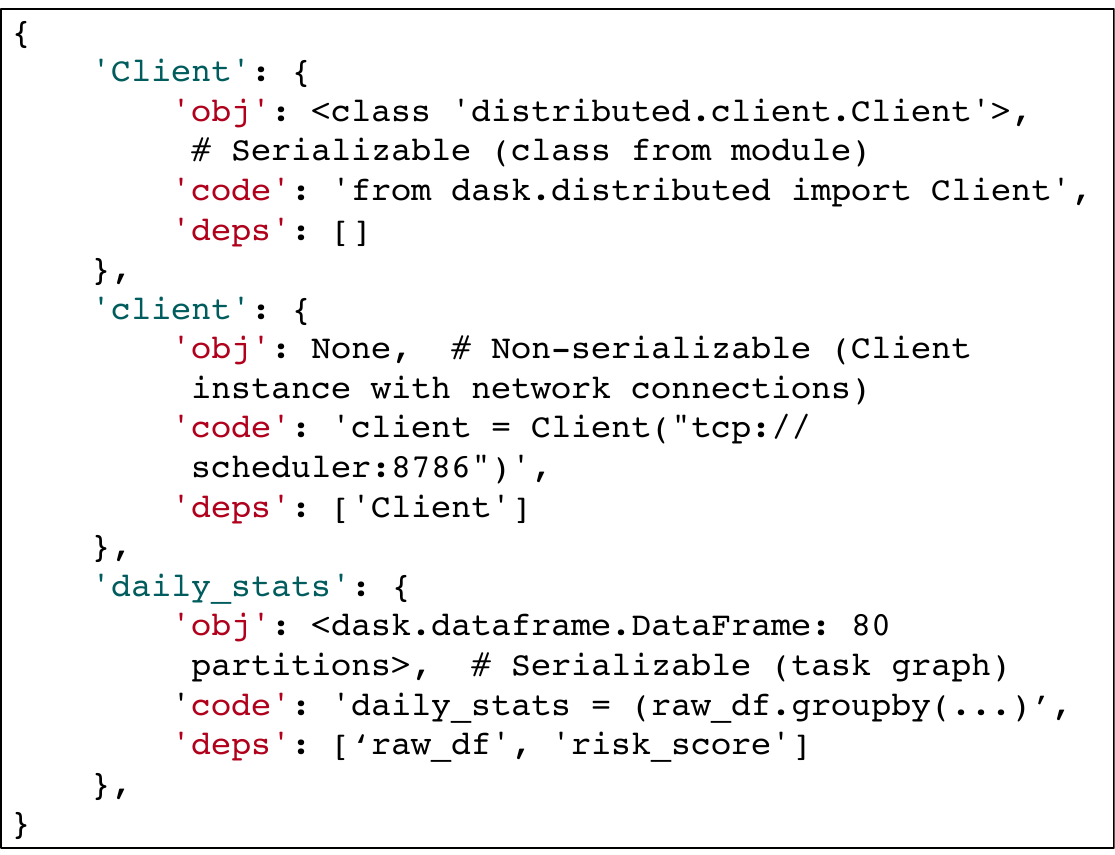}
  \caption{The schema of the checkpoint created by \toolname includes the name, code, and dependencies of each persisted object.}
  \label{fig:checkpoint-schema}
\end{figure}

To illustrate incremental checkpointing and shared-variable inference, Figure~\ref{fig:nbrewind-workflow} presents a running example with two notebook cells. The second row of the figure shows the variables identified for each cell via abstract syntax tree (AST) parsing. This static analysis conservatively captures all variables declared or referenced within a cell. For example, in Cell~1 the parsed variables are \texttt{[Client, client, ddf, df]}, while Cell~2 yields a different set of variables.

Using 
the reverse index of memory locations (shown in the last row of Figure~\ref{fig:nbrewind-workflow} for Cell~1), \toolname identifies shared variables, i.e., variables that reference the same underlying runtime object. For instance, the variable $raw\_df$ shares a memory location with $df$ due to the assignment in Line~7 of Cell~1 (shown in the third row of Figure~\ref{fig:nbrewind-workflow}), and is therefore marked as a shared variable.

In contrast, in Cell~2, although $state\_ref$ is initially assigned to $daily\_stats$, they are not marked as shared variables. This is because $daily\_stats$ is subsequently reassigned in Line~4 of Cell~2, breaking the shared memory relationship. Since shared-variable determination is performed at cell-level granularity, the position of declarative and assignment statements affects the inferred sharing. For example, if Line~4 of Cell~2 were moved to a subsequent cell, then within Cell~2, \toolname would infer $statef\_ref$ and $daily\_stats$ as shared variables. However, \toolname updates shared-variable information immediately upon reassignment, ensuring correctness of incremental checkpoints.

Checkpointing decisions are based on the union of variables identified via AST parsing and those inferred through shared-variable tracking. As a result, in Cell~2, even though $raw\_df$ is the only variable parsed from the AST, \toolname also checkpoints $df$, since $raw\_df$ is shared with $df$ through prior assignments.

Figure~\ref{fig:checkpoint-schema} further shows the metadata checkpointed for each variable at each cell. This metadata includes: (i) the serialized Python object when possible, (ii) the code that produced the object, and (iii) any environment or execution dependencies required for reconstruction. For example, \texttt{Client} is a Python class that can be serialized directly and has no prior code dependencies. In contrast, the \texttt{client} object represents a live Dask connection and cannot be serialized; such objects are not checkpointed. Instead, \toolname reconstructs them during replay by re-executing the relevant initialization code. The variable $daily\_stats$, a Dask DataFrame, is serializable, and its checkpoint includes both the object and the code expression that produced it, along with its transitive dependencies.

Finally, \toolname performs checkpoint deduplication across cells. As shown in Figure~\ref{fig:nbrewind-workflow}, several variables (e.g., $df$ and $raw\_df$) appear in multiple checkpoints. Although these variables are included in successive checkpoints due to the possibility of change, their serialized representations may be identical. \toolname detects and eliminates duplicate checkpoint entries across cells, significantly reducing storage overhead, as demonstrated in our experimental evaluation.

\section{Partial Re-Execution}
\label{sec:Partial Re-Execution}

\subsection{Stateless Distributed Workflows}


Consider the Dask-based climate analysis workflow shown in Figure~\ref{fig:notebook-example}. Figure~\ref{fig:task-dag} illustrates an abstracted version of its task dependency graph, where Dask partitions the dataset into 3 chunks for parallel processing. The workflow spawns tasks
that (1) load temperature data from a Parquet file (T1--T3), followed by tasks to (2) filter nulls values (T4--T6), and (3) computes anomalies (T7--T9). Then it performs (4) parallel aggregation by region and day (T10--T12) and (6) a reduce operation to persist the intermediate aggregated result (T13). It then runs tasks to (7) map partitions that apply the risk score function (T14--T16), and finally (8) a reduce operation that persists the results (T17).

During iterative development, users frequently modify such workflows. Adding a new weather station file creates a new parallel branch while leaving T1 through T12 unchanged. Modifying the risk score function affects only T14, T15, T16, T17 while leaving T1 through T13 unmodified. However, standard execution with Dask and DaskVine forces complete re-execution because neither framework maintains persistent state across runs. DaskVine acts as a stateless executor that submits tasks to TaskVine without awareness of previous executions, and TaskVine does not track task history or cache outputs between workflow invocations. When modifying the risk score function, only 3 of 16 tasks require re-execution, yet the stateless model re-executes all 16 tasks.

\subsection{Stateful Execution with \textit{RewindManager}}

We introduce a persistence layer that transforms TaskVine into a stateful workflow manager. The architecture adds RewindManager, which wraps the standard TaskVine manager and interposes on task submission and completion. The execution flow becomes: Dask $\rightarrow$ DaskVine $\rightarrow$ RewindManager $\rightarrow$ TaskVine. RewindManager maintains a transaction log recording task fingerprints and execution metadata, plus a task cache storing output files.

When a task is submitted, RewindManager computes a fingerprint capturing the task's code and inputs, then queries the transaction log. On cache hit, it retrieves cached output files and returns a synthetic task result without executing it. On cache miss, the task proceeds to TaskVine for execution, and upon completion, the fingerprint and outputs are logged to the cache. Returning to Figure~\ref{fig:task-dag}, when modifying the risk score function, tasks T1 through T13 produce identical fingerprints and hit the cache, while T14, T15, T16, 17 have changed fingerprints and execute with the modified function. When adding a new weather station file, tasks T1 through T9 hit the cache, the new branch executes, and downstream aggregation tasks re-execute because their input fingerprints change.

\subsection{Task Fingerprinting Mechanism}

Task fingerprinting captures all elements including computational logic and input dependencies influencing task output. The strategy differs between command-based and Python function tasks.

\noindent \textbf{Command-Based Tasks.} The fingerprint hashes the command string concatenated with SHA-256 content hashes of all declared input files. For each input, the system reads file contents and computes a hash. The complete representation is serialized to JSON with sorted keys and hashed with SHA-256. This provides strong guarantees when all inputs are declared. 

\begin{figure}
  \centering
  \includegraphics[width=\linewidth]{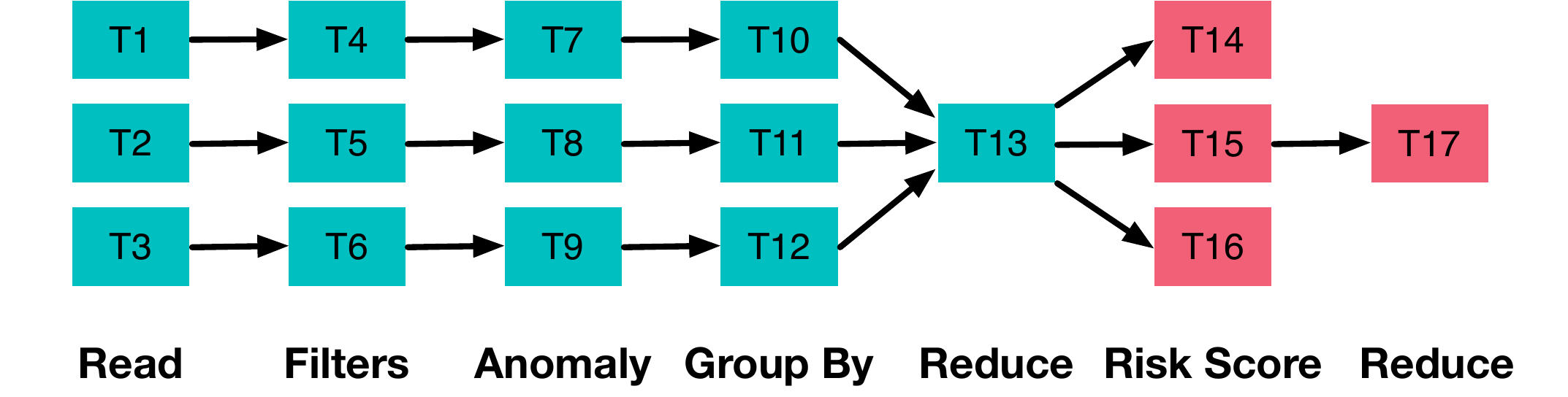}
  \caption{Partial execution of task graph from example shown in Figure~\ref{fig:notebook-example}. Update in `risk\_score` re-executes tasks shown in red and returns cached results for unchanged tasks shown in green.}
  \label{fig:task-dag}
\end{figure}

\noindent \textbf{Python Function Tasks.} Fingerprinting proceeds in two stages. First, a core hash captures function logic and arguments independent of input files. Arguments are canonicalized to remove non-deterministic UUID suffixes that Dask injects (e.g., \texttt{subgraphcallable-a1b2c3d4} normalizes to \texttt{subgraphcallable}). The system identifies user-defined functions in arguments, extracts their source code via introspection, and includes source code hashes. In Figure~\ref{fig:task-dag}, tasks T14, T15, T16 include the \texttt{risk\_score} function as an argument; modifying this function changes the source code hash, invalidating cached results. Arguments are serialized and hashed as well. The core hash combines hashed arguments, and user function source hashes. 

Second, the core hash is combined with input file hashes. Similar to command based tasks, for each user-declared input file, the system computes a SHA-256 hash. Input hashes are sorted and appended to the core hash to compute the final fingerprint.

\subsection{Fingerprinting Challenges}

\noindent \textbf{Undeclared Dependencies.} Shell commands may reference files not declared as inputs assuming a shared file system (e.g., \texttt{bash process.sh data/*.csv} depends on both CSV files and \texttt{process.sh}). If \texttt{process.sh} is not declared, its modifications are invisible to fingerprinting, causing incorrect cache hits.

\noindent \textbf{Non-Determinism} 
The system assumes deterministic functions where identical inputs produce identical outputs. Functions depending on external state (timestamps, random numbers) as part of the closure violate this assumption, causing cached outputs to diverge from fresh execution. For example when you de-serialize a Dask generated task function its byte representation will be different in each run even though the task has not changed because it might reference some workflow identifiers that are non-deterministic.

Dask generates unique identifiers for intermediate results (\texttt{finalize-abc123}, \texttt{subgraphcallable-xyz789}) that appear in task keys and arguments. Canonicalization strips the known UUID formats through pattern matching, though this is framework specific and fragile and unrecognized formats may cause unnecessary cache misses.

\section{Experiments and Evaluation}
\label{sec:Experiments and Evaluation}

We evaluate \toolname by applying it to real-world scientific notebook workflows that contain distributed computations and benefit from re-execution optimization. Our evaluation demonstrates that \toolname can significantly reduce notebook re-execution time through incremental checkpointing and partial re-execution of distributed tasks, while maintaining reproducibility guarantees. We assess \toolname's effectiveness across three key dimensions: (1) checkpoint creation overhead during audit mode, (2) time savings during repeat execution with unchanged notebooks, and (3) selective task re-execution efficiency when notebooks are modified.

\subsection{Experimental Setup}

\textbf{Computing Infrastructure.} All experiments were conducted on a SLRUM cluster with 4 compute nodes connected via a high-speed network. Each node was equipped with dual Intel Xeon Platinum 8259CL CPUs running at 2.50GHz and 4GB of memory. The cluster used NFS4 for shared file system access and XFS for local storage. Worker processes were configured with a minimum of 1 and maximum of 2 processes per node, allowing TaskVine to dynamically allocate resources based on workload demands.


\textbf{Dataset Description.} Our evaluation dataset comprises 5 notebooks from diverse scientific domains, each demonstrating different computational characteristics, data dependencies, and distributed execution patterns. Table~\ref{tab:notebook_dataset} provides a detailed characteristics of each notebook.

\begin{table*}
\centering
\caption{Characteristics and description of the five notebooks in the dataset.}
\label{tab:notebook_dataset}
\begin{tabular}{|l|c|c|c|c|c|p{6cm}|}
\hline
\textbf{Notebook} & \textbf{Cells} & \textbf{Unique Variables} & \textbf{Shared Variables} & \textbf{Dataset Size} & \textbf{Distributed Tasks} & \textbf{Description} \\
\hline
DV5 & 11 & 60 & 7 & 952 KB & 4 & High energy physics analysis performing distributed event processing on CMS detector data using Dask and Coffea frameworks. \\
\hline
MapReduce & 8 & 70 & 10 & 768 KB & 13 & Distributed word count pipeline implementing map-reduce pattern. Creates 12 parallel map and combine tasks for word counting and 1 reduce task for aggregation. \\
\hline
DConv & 7 & 27 & 5 & 16.4 MB & 1024 & Tile based image convolution applying kernel functions to partitioned image tiles using Python Imaging Library. Processes 1024 tiles with 2 different kernels. \\
\hline
RAG & 14 & 54 & 15 & 6.2 MB & 4 & Document chunking and indexing pipeline for BM25 based retrieval. Splits books into semantic chunks and builds inverted index for similarity search. \\
\hline
CTrend & 12 & 80 & 7 & 27.44 MB & 120 & Climate trend analysis processing temperature data from NOAA weather stations. Performs data cleaning, anomaly calculation relative to baseline period, and trend aggregation with statistical analysis. \\
\hline
\end{tabular}
\end{table*}

\subsection{Methodology}

Our experimental methodology is designed to evaluate \toolname across realistic notebook development and sharing scenarios. We compare notebook execution under 5 distinct conditions organized into three categories furthemore modifications to each notebook are explained later in this section:

\textbf{Dataset Description.} Our evaluation dataset comprises 5 notebooks from diverse scientific domains, each demonstrating different computational characteristics, data dependencies, and distributed execution patterns. Table~\ref{tab:notebook_dataset} provides a detailed characteristics of each notebook.

\noindent\textbf{Baseline Execution (Without \toolname):}
\begin{enumerate}
    \item \textbf{Initial Run:} Execute the original notebook from top to bottom using the standard Python kernel without any checkpointing or task caching. This represents typical first-time execution and establishes the performance baseline.

    \item \textbf{Modified Run:} Execute a modified version of the notebook using the standard Python kernel. Modifications include either adding new input data files or changing workflow code. This simulates iterative development where users make incremental changes and must re-execute the entire workflow.
\end{enumerate}

\noindent\textbf{Audit Phase (With \toolname Checkpointing):}
\begin{enumerate}
    \item \textbf{Initial Run:} Execute the notebook using \toolname's audit kernel with incremental checkpointing enabled. The kernel creates checkpoints after each cell execution, persisting only variables and objects that have changed since the previous cell or they were used in this cell. \toolname's TaskVine manager simultaneously logs all completed tasks with their input fingerprints and caches task outputs.

\end{enumerate}

\noindent\textbf{Repeat Phase (With \toolname Restoration):}
\begin{enumerate}
    \item \textbf{Repeat Without Change:} Execute the notebook using \toolname's repeat kernel without any modifications. The kernel restores cell outputs from checkpoints for unchanged cells and replays cached task results for all distributed tasks. This simulates sharing a notebook with a collaborator who wants to verify results without changes.

    \item \textbf{Repeat With Change:} Execute the modified notebook as used in Baseline modified run using \toolname's repeat kernel. The kernel identifies which cells and tasks are affected by the modifications and selectively re-executes only the invalidated portions. Unchanged cells are restored in-order from checkpoints, and unchanged tasks are replayed from the cache. This simulates the iterative development process.
\end{enumerate}

We add additional input data files or modify a task and expect the workflow to be executed partially. For each notebook, we also report the expected tasks to be replayed from the cache and re-executed. This expected behavior is confirmed from the results shown in Figure~\ref{fig:task_caching} and explained later in results. We apply the following modifications to each notebook for partial re-execution evaluation.
\begin{itemize}
\item \textbf{DV5:} Added 1 additional ROOT file containing events. Expected: 3 out of 4 tasks cached.
\item \textbf{MapReduce: }Modified reduce function to compute group counts in addition to total counts. Expected: 12 out of 13 tasks cached. 

\item \textbf{DConv:} Added 1 new kernel matrix (Gaussian filter). Expected: 1024 out 2048 tasks cached.
\item \textbf{RAG:} Added 1 additional book ($\sim$500KB text). Expected: 3 out of 4 tasks cached.
\item \textbf{CTrend:} Added 10 new weather station CSV files. Expected: 100 out of 121 tasks cached.
\end{itemize}


For each experimental condition, we measure:
\begin{itemize}
    \item \textbf{Execution time:} Total time from notebook execution start to completion, including all distributed task execution.
    \item \textbf{Task level cache statistics:} Number of tasks submitted, cached, and re-executed during repeat phase with modifications.
    \item \textbf{Storage overhead:} Size of checkpoints (after content-based de-duplication) and task cache (transaction log and intermediate files).
\end{itemize}

We assume that input data files do not change between successive runs, though new files may be added. Additionally, input data file names remain consistent between runs to enable accurate task fingerprinting and cache lookup. These assumptions align with common scientific workflow practices where input datasets are treated as immutable and new data is added as separate files. We also assume that the task functions do not contain any non deterministic values as arguments or in it's closure so that we can get a stable fingerprint of the task for effective caching.


\subsection{Results and Analysis}

\begin{figure*}[t]
  \centering
  \includegraphics[width=\textwidth]{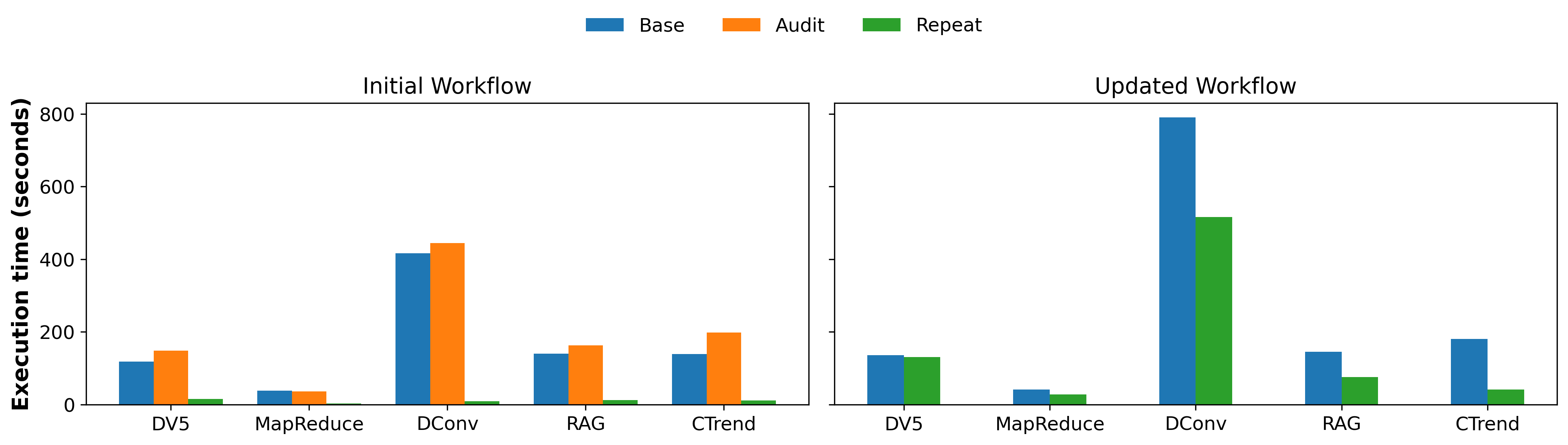}
  \caption{Execution time in seconds for notebooks for both audit and repeat phases compared with the baseline.}
  \label{fig:execution_times}
\end{figure*}

Figure~\ref{fig:execution_times} presents the execution times for each notebook under all experimental conditions. We observe that \toolname provides substantial benefits across all notebooks, with the magnitude depending on the notebook's workflow structure and modification patterns.

\noindent \textbf{Audit Overhead.} Audit mode incurs an average of  18\% overhead across notebooks ranging from 7-42\% : DV5 (25\%), MapReduce (1\%), DConv (7\%), RAG (16\%), and CTrend (42\%). Figure~\ref{fig:execution_times} shows these execution times. This overhead is attributed to caching the tasks and checkpointing the notebook state after the execution of each cell. This overhead becomes negligible when compared to the performance gains we achieve when repeating the notebook.

\noindent \textbf{Repeat Execution Without Modifications.} 
Repeat mode consists of (1) local kernel state reconstruction from incremental checkpoints, and (2) distributed task execution by replaying cached task results. This simulates sharing a notebook with a collaborator who wants to verify results without changes. Comparing repeat mode against baseline normal runs reveals significant speedup from eliminating redundant computation. \toolname achieves an average performance gain of 18.4x ranging from 8x to 46.2x: DV5 (118.29s $\rightarrow$ 15.69s), MapReduce (38.79s $\rightarrow$ 2.78s), DConv (416s $\rightarrow$ 9s), RAG (140s $\rightarrow$ 12s), and CTrend (139s $\rightarrow$ 11s) by restoring kernel checkpoints and replaying cached task outputs. Since DConv has the most time spent in distributed computation, it achieves the highest performance gains. These results reveal that using \toolname we can improve the speed of reproducing distributed workflow by order of magnitude.

\noindent \textbf{Repeat Execution With Modifications.} Comparing repeat execution with modifications against baseline modified runs demonstrates \toolname's effectiveness in executing partially updated workflows. The average performance gain is 2.08x, ranging from 1.04x to 4.41x depending on the modification scope: DV5 (136s $\rightarrow$ 130.50s), MapReduce (41.23s $\rightarrow$ 27.67s), DConv (791s $\rightarrow$ 516s), RAG (145s $\rightarrow$ 76s), and CTrend (181s $\rightarrow$ 41s). Since adding 10 new weather stations in CTrend requires processing only 20 new tasks while caching 100 existing tasks we gain the most time savings.

\begin{figure}[t]
  \centering
  \includegraphics[width=1\linewidth]{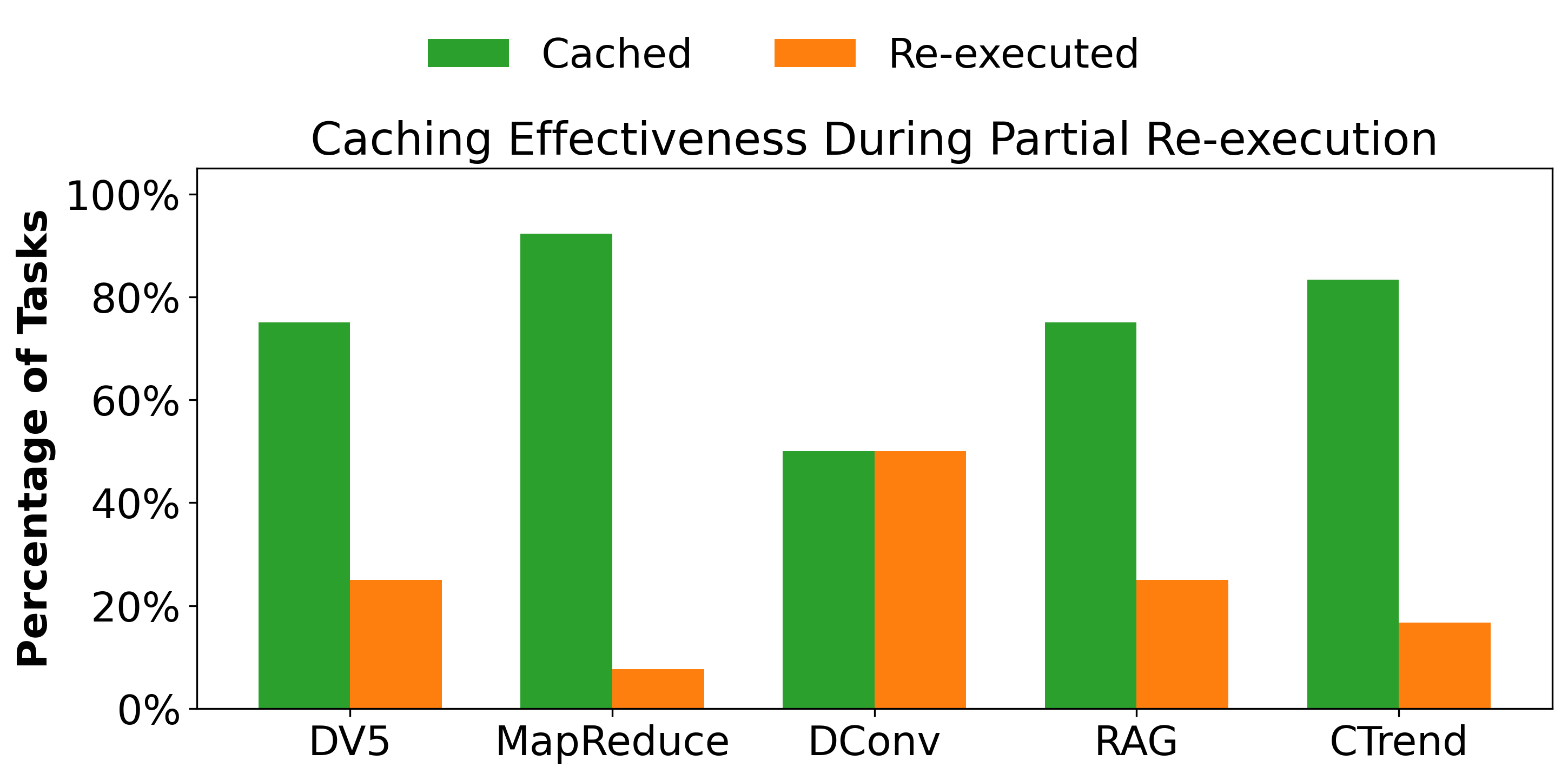}
  \caption{Percentage of Cached and Re-executed tasks in the updated workflow.}
  \label{fig:task_caching}
\end{figure}

\noindent \textbf{Task Caching Effectiveness.} The observed cache hit rates as shown in Figure-\ref{fig:task_caching} range from 50\% to 92.3\% across notebooks: DV5 (75\%, 3 of 4 tasks cached), MapReduce (92.3\%, 12 of 13 tasks cached), DConv (50\%, 1024 of 2048 tasks cached), RAG (75\%, 3 of 4 tasks cached), and CTrend (83.3\%, 100 of 120 tasks cached). The average hit rate for all notebooks is 75.1\%. This reveals that \toolname's effectiveness is determined by modification patterns and workflow structure. We identify two primary factors that govern caching behavior: adding new data versus modifying existing code, and the workflow DAG contains independent parallel tasks versus connected dependent tasks. 

For workflows with group of tasks that are processed independent of other tasks (DV5, RAG, CTrend), adding new input files creates a separate branch of tasks that leaves existing tasks unaffected. Adding 10 CSV files to CTrend, for instance, creates 20 new tasks while successfully caching the 100 existing tasks, achieving an 83.3\% hit rate. This pattern generalizes across workflows where each input file spawns independent processing pipelines.

Code modifications exhibit more complex behavior depending on modification location within the DAG. For example, modifying the reduce function in MapReduce re-executes only 1 of 13 tasks (92.3\% hit rate) because the change occurs at the workflow's terminal stage, leaving all upstream map tasks cacheable. In contrast, modifying early stage functions would cascade through all downstream tasks, eliminating caching benefits entirely. For instance, in DConv, updating the kernel matrix in convolution triggers a complete re-execution of workflow whereas extending the workflow with another kernel will achieve a 50\% cache hit.

\noindent \textbf{Storage Overhead.} Table~\ref{tab:storage_overhead} reveals that \toolname's storage costs decompose into two distinct components with different scaling behaviors: Python kernel state checkpoints and distributed task intermediate files. Incremental checkpointing with content-based de-duplication keeps kernel state storage minimal, ranging from 0\% to 57\% reduction in checkpoint size with an average reduction of 27.3\%. DConv achieves the highest de-duplication savings at 57\% (250 MB reduced to 108 MB) due to repeated references to large image arrays across multiple cells. The de-duplication mechanism effectively eliminates redundant data referenced by multiple variables and stores it only once. In contrast, intermediate file storage scales with the DAG structure. DConv's 1024 convolution tasks each produce an output image tile, generating substantial intermediate storage that persists for task cache replay. These intermediate files, while not directly visible to end users in the final notebook output, enable \toolname's selective re-execution capability by preserving task-level outputs.

\begin{table}[t]
\centering
\caption{Storage overhead breakdown showing incremental checkpoint sizes before and after content-based deduplication, and the intermediate task cache files.}
\label{tab:storage_overhead}
\begin{tabular}{|l|c|c|c|c|c|}
\hline
\textbf{Notebook} & \multicolumn{2}{c|}{\textbf{Checkpoint Size}}  & \textbf{Intermediate Files} \\
\cline{2-3}
        & \textbf{(Pre-Dedup)} & \textbf{(Post-Dedup)} & \textbf{}\\
\hline
DV5               & 420 KB  & 264 KB  & 112 KB \\
\hline
MapReduce         & 380 KB  & 332 KB  & 300 KB \\
\hline
DConv             & 250 MB  & 108 MB  & 159 MB \\
\hline
RAG               & 116 MB     & 81 MB     & 8.4 MB \\
\hline
CTrend            & 1.9 MB     & 1.9 MB     & 400 KB \\
\hline
\end{tabular}
\end{table}



\section{Related Work}
\label{sec:Related Work}

Distributed workflows in Jupyter notebooks have become quite popular in various scientific disciplines. They can also be used to efficiently execute large-scale multi-user deployments where notebooks launch distributed jobs in HPC environments~\cite{zonca2018deploying},\cite{hpc-jupyter},\cite{supercomputing-jupyter}. Similarly, Colonnelli et al.~\cite{colonnelli2022distributed} extend the Jupyter model so each cell maps to a workflow step that can run in cloud or on HPC resources. This has become faster and convenient, especially with the rise of systems like Floability~\cite{islam2025backpacks} which provide rapid, portable, and reproducible deployment of complex scientific workflows across a wide range of cyberinfrastructures. 

Intermediate state preservation has been shown to reduce redundant re-execution and enhance interactive performance in traditional HPC environments~\cite{arya2016design} and interactive notebook system~\cite{fang2025large}. Kishu~\cite{kishu} is a system for versioning notebook session states through incremental checkpointing. Li et al. propose ElasticNotebook~\cite{li2024demonstration} which enables live migration of notebook sessions across machines via lightweight state monitoring and graph-based optimization. Sato et al. \cite{sato2024multiverse} developed a computational notebook engine called Multiverse Notebook which allows user to time travel to any past state through cell-wise checkpointing using the POSIX fork to keep track of the state of cells, besides developing their own garbage collection strategy. Brown et al. focus on the relationships between cell and data dependencies in notebooks~\cite{brown2023facilitating}. 

While these efforts have made advances in capturing dependencies and guiding users through execution lineage, they typically either address portability, reproducibility, or provenance without explicitly targeting the optimization of notebook re-execution performance. Additionally, they do not monitor any distributed computation taking place in a notebook cell. \toolname distinguishes itself by unifying incremental checkpointing, content-based de-duplication, and application virtualization to explicitly optimize the performance of notebook re-execution for distributed workflows. 

For containerization, That et al.~\cite{fils2017sciunits} introduce Sciunit to create reusable research objects which uses a command-line tool to audit, store, and repeat program executions. ReproZip~\cite{chirigati2016reprozip} uses a similar approach to Sciunit and automatically track and bundles all dependencies of a computational experiment in a self-contained package. For reproducibility in notebook environments, Ahmad et al. build FLINC~\cite{ahmad2022reproducible} using Sciunit which creates reproducible notebook containers by virtualizing the user kernel through additional kernels for audit and repeat. 
Shankar et al. NBSLICER~\cite{shankar2022bolt} is another notebook-based tool that creates the forward and the backward program slices as a result of execution of each cell.

\section{Conclusion}
\label{sec:Conclusion}

With the increase in popularity of distributed workflows in notebooks, sharing them with collaborators and reproducing them in different compute environments remains a challenge. We introduce \toolname which improves the efficiency and reproducibility of distributed workflows within notebook environments by implementing a split-kernel architecture consisting of audit and repeat kernels. It enables users to save and restore the notebook state at cell-level by efficiently storing the notebook state using application level checkpointing. It tracks fine-grained state changes using a reverse index map of memory addresses to variables and 
uses content-based de-duplication for efficient space consumption. \toolname also integrates task-level memoization and logging in a distributed workflow through a stateful distributed workflow manager, allowing the system to selectively replay only those computations affected by code or data modifications. Using a notebook dataset of real-world case studies, we demonstrate that \toolname significantly reduces re-execution overhead while maintaining the portability required for cross-site collaboration on HPC systems.

In future, we will focus on expanding \toolname's capabilities to support a broader range of parallel and distributed computing systems and platforms. We will also perform improved integration with automated "backpack-like" packaging mechanisms for containerization and reproducibility to facilitate sharing with collaborators. We will also enhance the efficiency of \toolname to handle large-scale data-intensive distributed scientific computing workflows to facilitate seamless sharing of computational artifacts through integration with global data federations such as the National Science Data Fabric.


\bibliographystyle{IEEEtran}
\bibliography{references}

\end{document}